\documentclass[12pt,twoside]{article}
\usepackage{fleqn,espcrc1,epsfig}
\usepackage{amssymb}

\title{Effects of deformation in the three-body structure of $^{11}$Li}

\author{I. Brida, F.M.~Nunes and B.A. Brown
\address{ NSCL and Department of Physics and Astronomy
Michigan State University, East Lansing MI 48824 USA}
\thanks{This work is supported by by the National Science Foundation through grants PHY-0456656
and PHY-0244453, and the NSCL at Michigan State University.
}}

\begin{document}

\maketitle

\begin{abstract}
$^{11}$Li is studied within a three-body model $^9$Li$+n+n$ where the core is allowed to
be deformed and/or excite. In particular, we include reorientation couplings and couplings
between the two bound states of $^9$Li. Contrary to the other examples studied within this
model, we find that core excitation does not affect the structure of $^{11}$Li significantly.
Reorientation couplings of the deformed $^9$Li can change the ground state of $^{11}$Li from a
predominantly two neutron $(s_{1/2})^2$ configuration into a $(p_{1/2})^2$. In addition,
we see no evidence
for the existence of significant d-wave strength in its ground state, as opposed to the
prediction by shell model. A comparison with shell model is presented.
\end{abstract}

% ====================================================================

\section{Introduction}
\label{intro}

The prima donna of all halo nuclei is $^{11}$Li: it offers an exciting
experimental challenge but also questions the
basic concepts of a large variety of structure theories.
% summary of available measurements
Many experiments have been performed over the last decade, since the pioneering
total reaction cross section experiment at Berkeley \cite{kobayashi}. These include
Coulomb dissociation experiments \cite{ieki,zinser-2},
mass measurements \cite{young,gornov}, momentum distributions
of the $^9$Li coming from the two neutron removal \cite{hansen,orr,humbert}
and of $^{10}$Li coming from the one neutron removal \cite{zinser,simon},
proton inelastic scattering \cite{korsh}, as well as
beta decay studies \cite{borge,aoi,morrissey}.
We now know the system is bound by $295 \pm 35$ keV
against three-body
breakup \cite{young} with a momentum
distribution around $40$ MeV/c \cite{orr} (see {\it Addendum} at the end of the paper).
The ground state structure
contains two valence neutrons in $(s_{1/2})^2$ and $(p_{1/2})^2$ configurations
with approximately equal probability \cite{simon,borge}.
The r.m.s. matter radius of the system is $3.55 \pm 0.10$ fm \cite{radius}.
%As soon as we consider excited states the picture becomes less clear.

One cannot understand the Borromean nucleus $^{11}$Li without a good description of the
$n-core$ subsystem $^{10}$Li. Experimental programs are aware of this fact
but, as this nucleus is particle unbound, it presents a further technical obstacle.
The information on $^{10}$Li is summarized in \cite{tilley}.
Therein, it is possible to see the large number of experiments that
have been performed to measure $^{10}$Li's spectrum, but also the contradictory energy,
parity  and spin assignments made. Overall one can conclude the following:
(i)  the ground state of $^{10}$Li contains a valence neutron outside the $p_{3/2}$ subshell
in a $s_{1/2}$ state,  at approximately 50 keV or below;
(ii) there is a state around 0.4 MeV that can be understood as a neutron single particle
excitation into the $p_{1/2}$ orbital; (iii) there is no clear evidence for a d-state  below
3 MeV.

%summary of structure models
In the early days, three-body models of two neutrons and an inert $^9$Li core
were developed to describe properties of $^{11}$Li \cite{pairing,li11,fedorov95,esbensen}.
At that time nothing was known about $^{10}$Li, and theorists could play  the game
of adjusting freely the $n-^9$Li interaction in order to produce a sensible $^{11}$Li
ground state \cite{li11,fedorov95,garrido}. Models produced states where the two neutrons
were in $(2s_{1/2})^2$ and in $(1p_{1/2})^2$ states coupled to the ground state
of the core $^9$Li. The relative strength of these two components varied significantly
depending on the $n-core$ effective interaction. In \cite{garrido} a three-body force
was introduced in addition to the two-body $n-n$ and $n-core$ interactions.
In \cite{pairing,esbensen} a density dependent $n-n$ delta force  was used and
emphasis was given to the importance of pairing.
Three-body inert core models have been expanded to generate three-body continuum states
\cite{li11cont} and the complexity of these three-body scattering
states was analyzed within the context of proton inelastic scattering.

Microscopic calculations in the early nineties were unable to reproduce a realistic
binding energy for $^{11}$Li without artificially renormalizing the interactions,
namely forcing the $p_{1/2}$ orbital to be closer to the continuum
threshold (e.g. \cite{shell,koepf}). As pairing effects had been identified to be crucial \cite{pairing},
by introducing a phenomenological force in the pairing channel, a self consistent description
of the Li isotopic chain became possible within the relativistic Hartree-Bogolyubov
framework (e.g. \cite{meng}). Effective interactions valid near the driplines have meanwhile
been developed in shell model \cite{kara}. Nevertheless,
configuration mixing, required to produce a realistic ground state for $^{11}$Li,
is still introduced by hand. It is only recently that ab-initio methods have
reached mass A=11 \cite{gfmc,ncsm}. Although the general spectra look promising,
there is a clear limitation on the accuracy of the energy (and consequently
of the two neutron separation energy)
and much more effort is necessary to obtain reliable spectroscopic information.

%what are the open questions
There are still  some open questions regarding $^{11}$Li, even when considering the ground state
only.
%The first has to do with core excitation. To what extend is
%the inert core picture $^{11}$Li=$^9$Li+n+n a good approximation? And
Should the quadrupole deformation of $^9$Li play a role in the structure of $^{11}$Li?
%The second has to do with higher partial waves.
Assuming the low lying states of $^{10}$Li are
mostly neutron single particle excitations, then we should find a
d-wave around 2 MeV excitation energy, as in the case of $^{11}$Be. Even if experiment has been
inconclusive in this respect, if such a state were to exist,
what would be the implications for the ground state of $^{11}$Li? Would one
expect to also have  a strong $(d_{5/2})^2$ configuration in the ground state
of $^{11}$Li as in the ground state of $^{12}$Be?

On the issue of core deformation and/or excitation, $^{11}$Li is considered
within the Nilsson model \cite{kurath}. The core $^9$Li has an uncoupled proton
that can be either in a $K=3/2^-$ or $K=1/2^-$ orbital. In that simple picture,
the extra two neutrons in $^{11}$Li, occupy a different K-orbit and would not
be strongly affected by the orientation of the deformed axis.
In addition, a shell model calculation using two independent frequencies (one for the core, the other
for the halo neutrons) \cite{kuo} shows that core polarization due to the
halo neutrons is very small.

%about 3b model with core excitation
The three-body model with core excitation \cite{be11nunes,be12nunes1,be12nunes2}
allows us to address  these issues keeping all the three-body dynamics.
This model was applied
to the Be isotopes $^{11}$Be and $^{12}$Be.
The coupled channel description of the states in $^{11}$Be was produced
assuming a $^{10}$Be+n structure where the $2^+$ excited state of $^{10}$Be
was explicitly included in the model space. A rotational model was used for
the core but the strength of the coupling between the
two $^{10}$Be states was taken from the measured $B(E2)$ value. An effective $n+core$
interaction for $^{11}$Be could be made to reproduce
the parity inversion of the $1/2^+$ and $1/2^-$ bound states,
the spectroscopy of these states, as well as the d-wave resonance.
Based on such a realistic description of the $^{10}$Be+n subsystem \cite{be11nunes}
three-body calculations were performed using hyper-spherical coordinates and a
hyper-spherical harmonic basis \cite{be12nunes1}. In this formulation, the Faddeev equations
reduce to a set of hyper-radial coupled channel equations. Antisymmetrization was
taken into account in an approximate way by projecting out the lowest  $^{10}$Be+n  states
with large overlap with the occupied $1s_{1/2}$ and $1p_{3/2}$.
Our model was able to predict most features of the $^{12}$Be system \cite{be12nunes2}.
In particular, the ground state was predicted to contain roughly one
third of core excitation with neutrons in the $(d_{5/2})^2$.

If only the neutrons play a role in the low lying spectra
of $^{10}$Li and $^{11}$Li, one  expects similarities between the spectra of
these nuclei and those of $^{11}$Be and $^{12}$Be. It is then reasonable to think
of $^{10}$Li as $^{9}$Li+n and of $^{11}$Li as $^{9}$Li+n+n. The main difference between
the three-body model of $^{9}$Li+n+n and that of $^{10}$Be+n+n is the spin
of the ground state of the core. Exactly for this reason, computational needs for $^{11}$Li are much larger than for $^{12}$Be.  Also, the first excited state of $^{10}$Be ($2^+$) couples the valence neutrons to new orbitals whereas the ground state of $^9$Li $(3/2^-)$ enables a larger variety of orbitals than its excited state $1/2^-$. Note that, in $^9$Li,  there are reorientation couplings just for the ground state, which are not present in $^{10}$Be. Are these  differences important when modelling $^{11}$Li? Now that computational power is very much increased, it is timely to revisit the problem  under the constraints provided by all the new $^{10}$Li data.

% ====================================================================

\section{Ingredients for the calculations}

\begin{figure}[t!]
\centerline{
    \parbox[t]{.4\textwidth}{
\centerline{\psfig{figure= 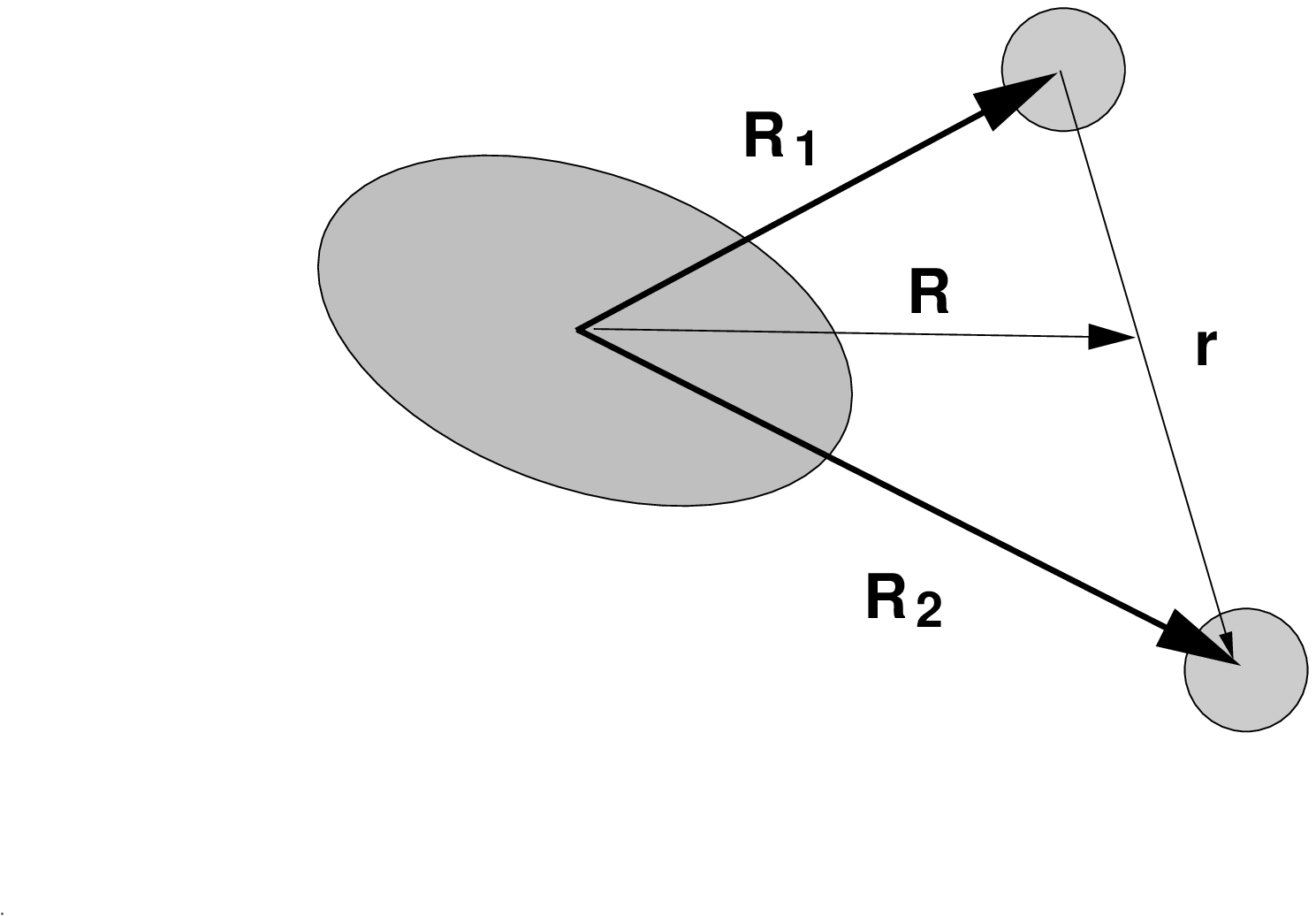,width=0.4\textwidth}}
\vspace{-1cm}
    \caption{Representation of the relevant coordinates in the model. }
\label{coord}}
}
\end{figure}
The Hamiltonian in the three-body model with core excitation can be written as:
\begin{equation}
H_{3B} = T_R + T_r + V_{nn}(\vec r) + V_{nc}(\vec{R_1},\xi) + V_{nc}(\vec{R_2},\xi) + h_{c}(\xi)\;
\label{h3b}
\end{equation}
where the coordinates are represented in Fig. \ref{coord} and $\xi$ is an internal
degree of freedom of the core.
These coordinates are transformed into the hyper-spherical coordinates
and through an appropriate expansion of the wavefunction,
the three-body Schr\"odinger equation is reduced to a set of coupled
channel equations dependent on hyper-radius $\rho$ and hyper-momentum $K$. The truncation of the basis
is done through the quantum number $K$.
Other details of the calculations can be found in \cite{be12nunes1,face}.

\begin{table}\begin{tabular}{|l|r|r|r|}\hline
 $J^{\pi}$    & $^{11}$Be  \\ \hline
$1/2^+$ &  0.0   \\
$1/2^-$ &  0.32  \\
$5/2^+(?)$ &  1.78  \\
$3/2^-(?)$ &  2.69  \\
\hline
\end{tabular}
\caption{\label{be11} Low lying spectrum of $^{11}$Be \cite{be11spect}. Excitation energies in MeV.}
\vspace{-0.5cm}
\end{table}
\begin{figure}[b!]
\centerline{
    \parbox[t]{.6\textwidth}{
\centerline{\psfig{figure=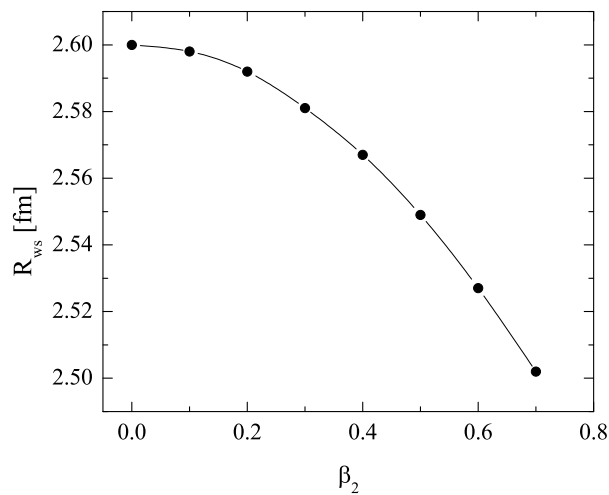,width=0.6\textwidth}}
\vspace{-1cm}
    \caption{Radii of the fitted n-$^{9}$Li interaction as a function
    of deformation. }
\label{rws-2b}}
}
\end{figure}
Here we discuss the physical inputs: the interactions and the model for the core.
The interaction between the two neutrons $V_{nn}$ is usually taken from a parameterization
of the low energy nucleon-nucleon phase shifts, which are well understood. We use
GPT interaction \cite{gpt}, neglecting the LL term. Most ambiguities reside in pinning
down the effective interaction between the core $^9$Li and each neutron. We will use the
notation $n l_{j} \otimes I$ to denote a specific $n-core$ configuration. For example
$1p_{1/2} \otimes 3/2^-$ corresponds to a valence $1p_{1/2}$ neutron coupled to the $3/2^-$ $^9$Li
ground state. A Woods-Saxon form plus spin-orbit is fitted to reproduce three main
structures in $^{10}$Li, namely the virtual state at $+50$ keV with configuration $2s_{1/2} \otimes 3/2^-$,
the resonance at $+400$ keV with configuration  $1p_{1/2} \otimes 3/2^-$ and a resonance at $+3.4$ MeV
with configuration  $1d_{5/2} \otimes 3/2^-$. Due to the lack of experimental data,
the position of d-resonance in $^{10}$Li is not known. By comparison with $^{11}$Be, with the
same number of neutrons, one would expect a resonance close to $2$ MeV excitation
(see Table \ref{be11}).
%The value $+3.4$ MeV was arbitrarily chosen because the  $^{10}$Li experimental spectra
%do no exhibit clear peaks above $\approx 1$ MeV.
We place it at $+3.4$ MeV:
if the d-wave resonance were at much higher energy, the binding
energy of $^{11}$Li would not be reproduced; if it were at much lower energy
it would become more bound than a p-wave resonance, for large deformation.
Also we consider that the $1p_{3/2}$ orbital is at $-4.1$ MeV,
to match the neutron separation energy of $^9$Li.

When there is deformation and core excitation, $l,j$ are no longer good
quantum numbers, and the nuclear state $J^{\pi}$ contains a superposition of different
$nl_{j} \otimes I$ configurations. Nevertheless, and for simplicity, we will refer
to any multi-component state $J^{\pi}$, by the component into which the state
collapses in the limit of no couplings.

\begin{figure}[t!]
\centerline{
    \parbox[t]{.6\textwidth}{
\centerline{\psfig{figure=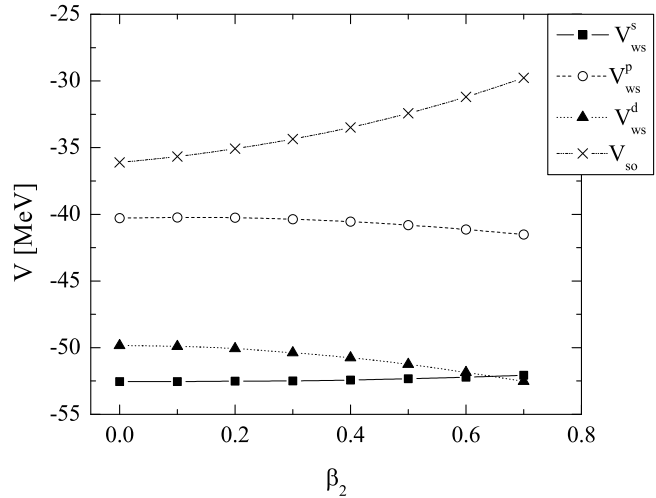,width=0.6\textwidth}}
\vspace{-1cm}
    \caption{Depths of the fitted n-$^{9}$Li interaction as a function
    of deformation.}
\label{vws-2b}}
}
\end{figure}
\begin{figure}
\centerline{
    \parbox[b]{.6\textwidth}{
\centerline{\psfig{figure=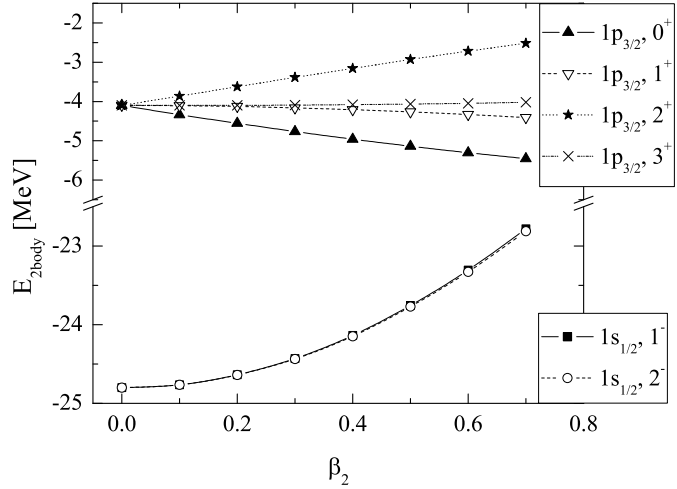,width=0.6\textwidth}}
\vspace{-1cm}
    \caption{Deeply bound states in $^{10}$Li as a function of deformation.
    The legend refers to $nl_j$ valence neutron orbitals
    (coupled to the $3/2^-$ ground state of $^9$Li) and total spin of $^{10}$Li.}
\label{li10a}}
}
\end{figure}
\begin{figure}[hb!]
\centerline{
    \parbox[t]{.6\textwidth}{
\centerline{\psfig{figure=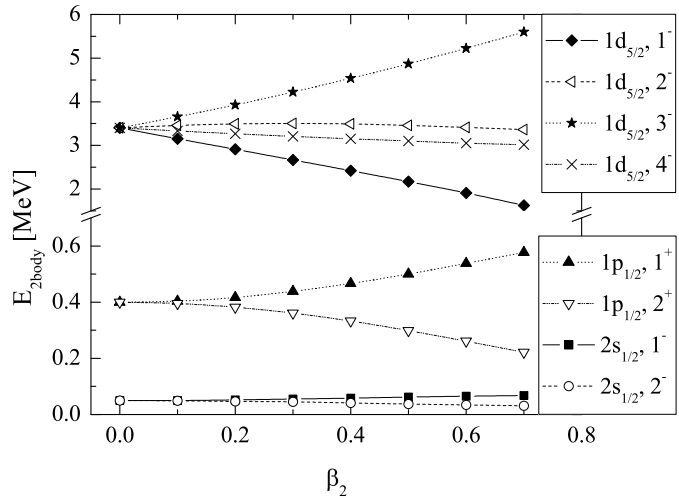,width=0.6\textwidth}}
\vspace{-1cm}
    \caption{Continuum states in $^{10}$Li, as a function of deformation. The legend refers to
     $nl_j$ valence neutron orbitals (coupled to the $3/2^-$ ground state of $^9$Li)
      and total spin of $^{10}$Li. Both s-wave states refer to virtual states while
    the other states correspond to real resonances.}
\label{li10b}}
}
\end{figure}

The form for the $n-core$ interaction is:
\begin{equation} \label{Vnc}
V_{nc}(\vec{R_i},\xi) =
\frac{V_{ws}^l}{1+e^{\left(\frac{R_i-R(\theta,\phi)}{a_{ws}}\right)
}} - \left(\frac{\hbar}{m_{\pi}c}\right)^2 (2 \vec l \cdot \vec s)
\frac{V_{so}}{4 R_i}\:\:\frac{d}{dR_i}
\left[1+e^{\left(\frac{R_i-R_{so}}{a_{so}} \right)} \right]^{-1},
\;\;\; i=1,2.
\end{equation}
Quadrupole deformation $\beta_2$ serves as the collective degree of freedom for the core
introduced in Eq.\ref{h3b}. The central Woods-Saxon part depends on the deformation through the radius
$R(\theta,\phi)  =  R_{ws} ( 1 + \beta_2\: Y_{20} (\theta,\phi) )$, where $\theta$ and $\phi$
are spherical angles in a rest frame of the core. The spin-orbit term is left undeformed.
In addition, the core is treated within a rotational model, parameterized in terms of $\beta_2$.
This is an oversimplification of the structure of $^9$Li, and thus we will not impose that
the strength of the couplings be determined by the $B(E2)$, which in any case is not known.
Volume conservation, however, is imposed on the central $n-core$ interaction \cite{be11nunes},
so when $\beta_2 \neq 0$, the radius parameter $R_{ws}$ is adjusted accordingly. We take
a standard value $R_{ws}=2.60$ fm when $\beta_2 =0$ and its dependence for $\beta_2 \neq 0$
is shown in Fig. \ref{rws-2b}. Radius $R_{so}$ was made equal to $R_{ws}$ at any deformation.
The corresponding diffusenesses are fixed to the standard value $a_{ws}=a_{so}=0.65$ fm.

The depth of the central part of the interaction between the core and the neutron depends on their
relative angular momentum $l$. We consider $V_{ws}^s$ and $V_{ws}^p$ for $l=0$ and $1$,
respectively  and the same $V_{ws}^d$ for all partial waves with $l\geq2$. The depth
$V_{so}$ of the spin-orbit term is $l$-independent. As mentioned before, the depths of $n-core$
potentials were fitted in order to reproduce the low lying continuum structure of $^{10}$Li
as well as the position of bound $1p_{3/2}$ neutron orbital.

In principle, due to the low excitation energy in $^9$Li (2.69 MeV), we should include
the $1/2^-$ state in our core model. Such calculations were performed but
results were not sensitive to the inclusion of this additional state. As mentioned in
Section \ref{intro} and due to its low spin value, there are no new orbitals brought
into the configuration space when including this state. For this reason in $^{11}$Li
reorientation is significant but core excitation is not. Note that the model space
when both core states ($3/2^-$;$1/2^-$) are included increases considerably.
We will present results for the case where only the ground state $3/2^-$ of $^9$Li
is taken into account.

In addition, the deformation parameters of $^9$Li can also take negative values
(if the core is oblate). We found that the quadrupole force for oblate shapes,
produces more repulsion when compared to the prolate case, and therefore less binding energy.
Based on the known positive quadrupole moment of $^7$Li, we expect this to also be
positive for $^9$Li. Therefore, we consider always positive deformation for
all results discussed in this work.

In case of an undeformed $^9$Li, all $J^\pi$ states in $^{10}$Li originating from a given $nl_j$ neutron coupled
to the core are degenerate. The degeneracy is, however,
removed as soon as the spherical symmetry is broken by non-zero deformation. In the deformed cases,
fitting potentials means to adjust their depths such that the centroids of $1p_{3/2}$,
$2s_{1/2}$, $1p_{1/2}$ and $1d_{5/2}$ orbitals are kept at $-4.1$ MeV, $+50$ keV,
$+400$ keV and $+3.4$ MeV, respectively. Again here, the $1d_{5/2}$ centroid at $3.4$ MeV ensures
that even when $\beta_2=0.7$ the splitting does not produce inversion of the p-state
and the d-state. The variation of potential depths with deformation
is shown in Fig. \ref{vws-2b}. As in $^{11}$Be, the spin orbit strength needed is large
\cite{be11nunes}. The corresponding two-body bound states and the
lowest resonances in $^{10}$Li
are shown in Fig. \ref{li10a} and Fig. \ref{li10b}, respectively.

Finally, in order to approximately satisfy the exclusion principle, the bound $1s_{1/2}$ and
$1p_{3/2}$ neutron orbitals are projected out of the model space before diagonalization
\cite{be12nunes1}.
%

% ====================================================================

\section{Results as a function of deformation}

\begin{figure}[t!]
\centerline{
    \parbox[t]{.5\textwidth}{
\centerline{\psfig{figure=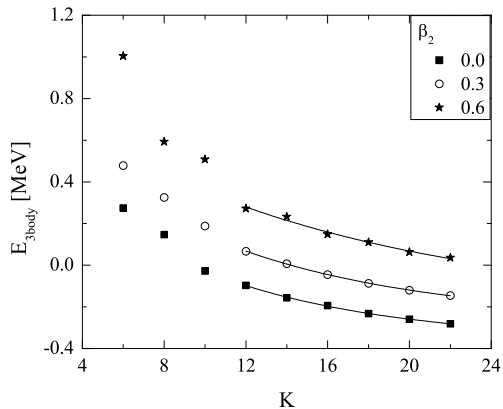,
width=0.5\textwidth}}
\vspace{-1cm}
    \caption{Convergence behaviour for $^{11}$Li as a function
    of the hyper-momentum K. The lines are an exponential fit.}
\label{conv}}
}
\end{figure}
%
%First we consider $^{11}$Li built on  the $3/2^-$ ground state of  $^9$Li only,
%and introduce quadrupole reorientation effects.
Calculations were performed using
EFADD \cite{efadd}. Our model space consists of 18 Laguerre polynomials for the hyper-radial part,
reaching out to $\rho_{max}=20$ fm, and 30 Jacobi polynomials for the hyper-angular part.
Calculations were performed up to $K_{max}=22$.

In Fig. \ref{conv} we show the convergence of the three-body binding energy of $^{11}$Li,
for $\beta_2=0.0, 0.3$ and $0.6$, as a function of hyper-momentum. On one hand, the
convergence exhibits the well known exponential dependence when $K_{max} \gtrsim 12$
for all deformations studied. On the other hand, in all cases, the convergence rate
is very slow, much slower than in the $^{12}$Be case (see Fig. 6  of \cite{be12nunes1}).
Moreover, the convergence rate decreases with increasing deformation.
It is thus necessary to use the extrapolated energy value for an
accurate binding energy.
We show results up to $\beta_2 = 0.7$ (this value is already unrealistically large).

\begin{figure}[t!]
\centerline{
    \parbox[t]{.5\textwidth}{
\centerline{\psfig{figure=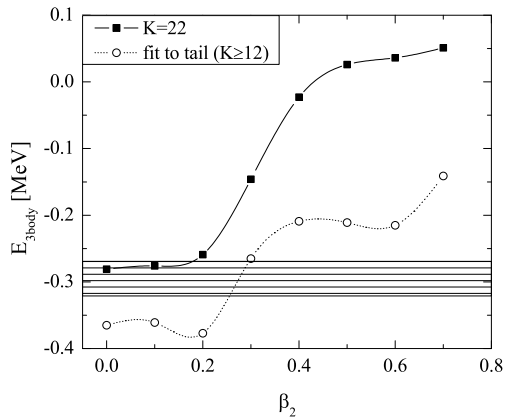,
width=0.5\textwidth}}
\vspace{-1cm}
    \caption{Three-body energy of the ground state of $^{11}$Li as a function
    of  deformation. Shaded region corresponds to data \cite{young}}
\label{li11energy}}
}
\end{figure}
\begin{figure}[t!]
\centerline{
    \parbox[t]{.5\textwidth}{
\centerline{\psfig{figure=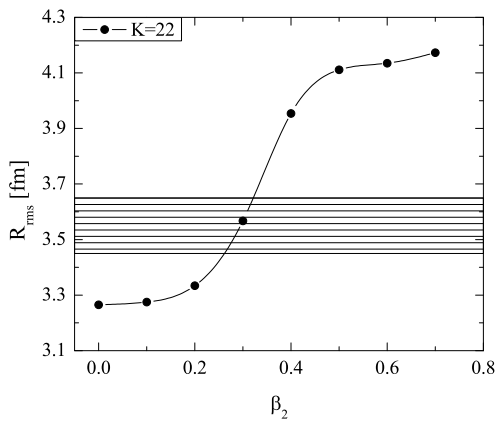,
width=0.5\textwidth}}
\vspace{-1cm}
    \caption{Radius of $^{11}$Li as a function
    of  deformation. Shaded region corresponds to the value consistent
    with reaction data \cite{radius}}
\label{li11rad}}
}
\end{figure}
Results for the three-body energy of $^{11}$Li are presented in Fig. \ref{li11energy}.
We plot both the values for $K_{max}=22$ and those obtained through the extrapolation in K.
Contrary to the case of $^{12}$Be where the energy gain was large,
here a very small additional binding is obtained from the
quadrupole coupling, and instead, as the deformation becomes large,
the system becomes less bound.  The r.m.s. radii of the corresponding wavefunctions are
shown in Fig. \ref{li11rad}.
Here we have used $2.32$ fm for the r.m.s. radius ($R_{rms}$) of the $^9$Li core  \cite{li9radius}.
These radii were obtained with $K_{max}=22$, but the $R_{rms}$ variation between $K_{max}=20$
and $K_{max}=22$ is less than $1\%$.
Our prediction for the charged radius obtained with $\beta_2=0.3$ is $R_{ch}=2.49$ fm, in agreement
with the very recent measurement \cite{sanchez}.
The experimental two-neutron separation energy and
the r.m.s. matter radius impose a constraint on $\beta_2 \lesssim 0.3$.
Note that the extrapolation was only done for extracting converged energies,
since it is the only observable that has a well established exponential dependence on  $K_{max}$.

\begin{figure}[t!]
\centerline{
    \parbox[t]{.5\textwidth}{
\centerline{\psfig{figure=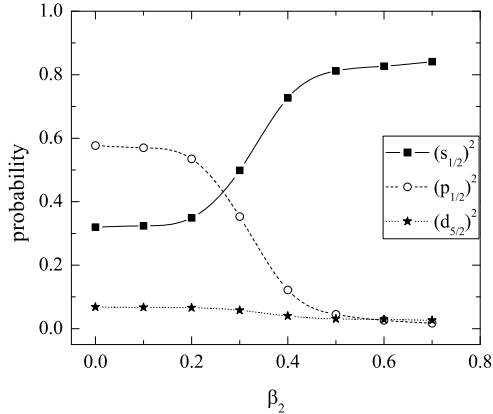,
width=0.5\textwidth}}
\vspace{-1cm}
    \caption{Structure of the ground state of  $^{11}$Li as a function
    of deformation. }
\label{li11prob}}
}
\end{figure}
We now look at the structure of the state. In Fig. \ref{li11prob} we show the probabilities
associated with the three main components of the system, namely $(s_{1/2})^2$, $(p_{1/2})^2$
and $(d_{5/2})^2$. If $\beta_2$ is small, the ground state of $^{11}$Li is almost $60$\%
$(p_{1/2})^2$; for large $\beta_2$, it becomes  more than $80$\% $(s_{1/2})^2$. The region
around $\beta_2=0.3$ corresponds exactly
to the transition between these two configurations and both are populated with equal
probability. Regardless of the deformation, the $(d_{5/2})^2$ configuration is always small ($< 7$\%).
This result is in contrast with the three-body calculations for $^{12}$Be in \cite{be12nunes2}
where the $(d_{5/2})^2$ configuration was $\approx 30$\%.

Finally, one has to realize that there is no unique parameterization for
the effective $n-core$ interaction. We have convinced ourselves that the
features  shown here for the structure of this three-body system, do not
result from a specific parameterization. Other parameterizations, using
other radii or spin-orbit parameters, produce exactly the same features.
These emerge from the constraints of continuum states in $^{10}$Li.
Variations on these constraints are considered at the end of the next section.

% ====================================================================

\section{Comparison with shell model}

One may expect $^{11}$Li to have a similar structure to $^{12}$Be.
Three-body calculations for $^{12}$Be have been compared with data
and shell model in the p-sd shell using the WBP interaction.
The results for the first low lying states of $^{12}$Be from \cite{be12nunes2} are presented in
Table \ref{be12}. As can be seen, both models are able to reproduce the first $0^+$, $2^+$ and $1^-$
states very well. Not shown, the two models also agree on the structure of the ground state of $^{12}$Be.
A three-body versus shell-model comparison of spectra for $^{11}$Li is not easy because $^{11}$Li excited states
are two-particle unbound.
\begin{table}\begin{tabular}{|l|r|r|r|}\hline
 $J^{\pi}$    & 3-body & shell model & experiment \\ \hline
$0^+$ &  0.0 & 0.0 & 0.0   \\
$2^+$ &  2.1 & 1.82 & 2.11 \\
$0^+$ &  2.2 & 0.37 & 2.24 \\
$1^-$ &  2.6 & 2.62 & 2.73 \\
\hline
\end{tabular}
\caption{\label{be12} Spectrum for $^{12}$Be: comparing three-body results
with shell model and experiment \cite{be12nunes2}. Excitation energies in MeV.}
\vspace{-0.5cm}
\end{table}

For the shell-model structure of $^{11}$Li we use the WBP Hamiltonian of \cite{sm}.
The WBP Hamiltonian is applicable to pure $N \hbar \omega$ model-space truncations
and is determined from a least squares fit of potential-model interaction strengths
to experimental states which are relatively pure $0 \hbar \omega$, $1 \hbar \omega$
and $2 \hbar \omega$ in structure. Application of this model to $^{12}$Be and $^{11}$Li
gives $0 \hbar \omega$ and $2 \hbar \omega$ states which are degenerate within 100 keV.
This implies wavefunctions for both  $^{12}$Be and $^{11}$Li which are
$50$\% $(s)^4(p)^{A-4}$ ($0 \hbar \omega$) plus
$50$\% $(s)^4(p)^{A-2}(sd)^2$ ($2 \hbar \omega$) and other small amplitudes.

In shell model, the $0 \hbar \omega$ $^{11}$Li wavefunction must have 2 neutrons in the $p_{1/2}$ orbit
while for the $0 \hbar \omega$ $^{9}$Li it predicts 0.79 neutrons in the $p_{1/2}$ orbit.
This means that $^9$Li is not a $p_{3/2}$ closed core, and consequently,
the $0 \hbar \omega$ $^{11}$Li wavefunction does not look like
$0 \hbar \omega$ $^{9}$Li  plus two $p_{1/2}$ neutrons.
The excitation energy of the $1/2^-$ $^9$Li state predicted by shell model is lower than
the experimental value ($1.395$ MeV to be compared with the experimental value of $2.69$ MeV).
The  $2 \hbar \omega$ $^{11}$Li wavefunction has the structure of
two-neutrons in the sd-shell outside the $^9$Li core, with occupancies
of 50\%, 29\% and 18\% for $s_{1/2}$, $d_{5/2}$ and $d_{3/2}$ respectively.

%Shell model predictions for $^{10}$Li and $^{11}$Li resemble those for $^{11}$Be and $^{12}$Be.
%Calculations  for $0 \hbar \omega$ and $2 \hbar \omega$ here presented use the WBP interaction.
%Since in shell model, these two configurations are degenerate, the mixing is 50-50.
The $0 \hbar \omega$ (negative parity) and $1 \hbar \omega$ (positive parity)
spectrum for $^{10}$Li with WBP, up to 4 MeV,  is presented in Table \ref{li10sm},
of which the
first doublet $2^-$,$1^-$ corresponds to $s_{1/2}$ and the second doublet $1^+$,$2^+$
to   $p_{1/2}$ strength. The lowest state coming from the
$d_{5/2}$-wave quadruplet is at $E_x(4^-)=2.05$ MeV, much lower than the energy
we considered in the three-body calculations ($3.4$ MeV).

\begin{table}\begin{tabular}{|c|r|}\hline
 $J^{\pi}$    & $E_x$ \\ \hline
$2^-$ &  0.00 \\
$1^+$ &  0.23 \\
$1^-$ &  1.11 \\
$2^+$ &  1.15 \\
$0^-$ &  1.73 \\
$0^+$ &  1.74 \\
$4^-$ &  2.05 \\
$1^-$ &  2.56 \\
$1^+$ &  3.14 \\
$2^-$ &  3.25 \\
$2^+$ &  3.29 \\
$1^-$ &  3.69 \\
$3^-$ &  3.74  \\ \hline
\end{tabular}
\caption{\label{li10sm} Shell model spectrum for $^{10}$Li. Excitation energies in MeV.}
\vspace{-0.5cm}
\end{table}
For the ground state of $^{11}$Li, shell model predicts 15\%  d-wave.
%However, the overlaps $\langle ^{11}$Li$(g.s.) | a^\dagger;1| ^{10}$Li (g.s)$ \rangle$ show
%three times more spectroscopic strength for the s-wave than for the $d_{5/2}$.
%In any case, shell model predicts a considerable amount of d-wave in the
%ground state of $^{11}$Li.
This $15$\% d-component in $^{11}$Li comes mainly from the low lying d-waves
present in the $^{10}$Li spectrum.

Also relevant for the discussion is the overlap
$\langle ^{11}$Li$(g.s.) | (a^\dagger, a^\dagger)_L| ^{9}$Li (g.s)$ \rangle$
which is predicted to be very strong and nearly exhausts all the spectroscopic strength.
This is a clear indication that the first excited state of $^9$Li is not important in the
description of the ground state of $^{11}$Li.

It is important to understand the implications of the $^{10}$Li structure on $^{11}$Li.
Albeit the large number of experiments, a close study of \cite{tilley} raises questions about
the precision with which states in $^{10}$Li are known. We have explored the possibility
of different assumptions for the states to which we fit the $n-core$ interaction, namely
$s_{1/2} \otimes 3/2^-$ at $+50$ keV; $p_{1/2} \otimes 3/2^-$ at $+400$ keV and
$d_{5/2} \otimes 3/2^-$ at $+3.4$ MeV. Of these, the least uncertain is the $^{10}$Li ground state.
We have checked that the main features of the present work are not changed by moving the
p-wave to $500$ keV. More important is the uncertainty in the location of the d-resonance.
There is no clear experimental evidence for a d-wave at $+3.4$ MeV, or at any lower energy. If it
is a broad resonance, or it is superposed by other states belonging to the doublets,
it could be much harder to observe experimentally.
The width of the d-wave resonance increases
rapidly with energy. If a $^{10}$Li d-state were at $1.2$ MeV as in $^{11}$Be, the width
would be $150$ keV. The $^{10}$Li $E_x=2.05$ MeV predicted by shell model has
$\Gamma=420$ keV and if it were shifted to $E_x=3.4$ MeV as in the three-body model,
the width would become $\Gamma=1.2$ MeV. Such a broad state would be very hard to measure.

As seen above, shell model predicts a $4^-$ d-wave
state at around 2 MeV. What happens in the three-body model if the d-wave is  pushed down?
We refitted the $n-core$ interaction with $\beta_2=0.3$, fixing the centroid of the
d-resonance at $2.5$ MeV. The quadruplet states ($1^-, 2^-, 3^-, 4^-$) move from
(2.67; 3.50; 4.22; 3.21) to (1.79; 2.53; 3.32; 2.36) (MeV), respectively.
An immediate consequence is the additional binding in the $^{11}$Li system ($\approx 150$ keV).
However this extra d-wave attraction is not sufficient to change the structure of
$^{11}$Li's ground state, which remains essentially
$(s_{1/2})^2$ and $(p_{1/2})^2$, with $(d_{5/2})^2$ only up to  $\approx 10$\%.

\section{Conclusions}

We perform three-body calculations for $^{11}$Li including deformation and excitation
of $^9$Li. We find that reorientation effects of the core can account for the known
configuration admixture of s-waves and p-waves in the ground state of $^{11}$Li.
With a simple rotational model for the core, it is possible to obtain
an effective interaction for $n-^9$Li, fitted to the properties of $^{10}$Li,
that produces the two neutron binding energy,
the r.m.s matter radius, the charged radius, and the structure of $^{11}$Li
all consistent with experiment.
Core excitation in the ground state of this three-body system
is found to be unimportant. This result is confirmed by shell model calculations.
Finally the d-wave strength is predicted to be very small ($\approx 7$\%)
in the three-body model. This is in disagreement with shell model predictions that
show roughly $30$\%, $50$\%, $20$\%  for $(s_{1/2})^2$, $(p_{1/2})^2$, and $d^2$ components.
The main reason for this discrepancy comes from the corresponding states in
the subsystem $^{10}$Li. So far, experiments have not been able to make a clear
statement about d-states
in $^{10}$Li. In our $n-core$ effective interaction we have assumed that
such states would be above 3 MeV. However, shell model produces
d-states that are at much lower energy (around 2 MeV), similar to $^{11}$Be.
Resolving experimentally the d-states in $^{10}$Li will
settle once and for all the  structure of the ground state of $^{11}$Li.
This experiment should be done with a reaction starting from $^9$Li rather than knock-out
from $^{11}$Li since there is not much d-wave in $^{11}$Li to start with.
One possibility would be to repeat the $^9$Li(d,p)$^{10}$Li \cite{jeppesen} with
higher beam energy.

\section*{\it Addendum}
After the completion of this work we learned about the recent high precision
mass measurement of $^{11}$Li: $S_{2n}=376 \pm 5$ keV \cite{bachelet}.
Even though the experimental accuracy is now well beyond the accuracy of the three
body model, this additional binding may be an indication that the d-wave is below
$3$ MeV.

\end{document}